\documentclass[prb,twocolumn,showpacs]{revtex4}
\usepackage{graphicx} 

\usepackage{amsmath,amsthm,amssymb,mathrsfs}
\usepackage{bm}

\begin{document}

\title{Proposal of an experimental scheme for determination of penetration depth of transverse spin current by a nonlocal spin valve}

\author{Tomohiro Taniguchi and Hiroshi Imamura}
\email{h-imamura@aist.go.jp}
 \affiliation{
 Spintronics Research Center, National Institute of Advanced Industrial Science and Technology, 
 Tsukuba, Ibaraki 305-8568, Japan
 }

 \date{\today} 
 \begin{abstract}
  {
  We theoretically propose an experiment to determine the penetration depth of a transverse spin current 
  using a nonlocal spin valve with three ferromagnetic (F) layers, 
  where the F${}_{1}$, F${}_{2}$, and F${}_{3}$ layers act as 
  the spin injector, detector, and absorber, respectively. 
  We show that the penetration depth can be evaluated 
  by measuring the dependence of the spin signal (magnetoresistance) 
  on the thickness of the F${}_{3}$ layer. 
  }
 \end{abstract}

 \maketitle



\section{Introduction}
\label{sec:Introduction}

There has been great deal of attention 
paid recently to the spin transport in nano-structured ferromagnetic materials 
because of their potential application to spintronics devices 
such as magnetic random access memory and microwave oscillators. 
The giant and tunnel magnetoresistance effects \cite{baibich88,binasch89,valet93,yuasa04,parkin04}
and spin torque effect \cite{slonczewski96,berger96} are key physical phenomena 
for the operation of these devices. 
The origin of these phenomena is the spin dependent electron transport, 
i.e., the spin current. 
The relaxation of the spin current in nonmagnetic (N) materials 
has been investigated 
using, for example, a nonlocal spin valve system 
\cite{johnson88,johnson93,johnson02,jedema01,jedema03,kimura06,kimura07a,kimura07b,kimura08}. 
In ferromagnetic (F) materials, 
on the other hand, 
we should distinguish the longitudinal and the transverse spin currents, 
whose spin polarizations are parallel and perpendicular to the local magnetization, 
and which are the origins of the magnetoresistance effect \cite{valet93,takahashi03} 
and the spin torque effect \cite{slonczewski96,berger96,brataas01}, respectively. 
Compared to the extensive studies on the spin relaxation of the longitudinal spin current \cite{valet93,fert99,takahashi03,godfrey06,bass07}, 
there have been very few studies on the spin relaxation of the transverse spin current.


The penetration depth of the transverse spin current (accumulation) is 
an important quantity 
characterizing its spatial spin relaxation 
\cite{taniguchi08a,taniguchi08d,taniguchi08b,taniguchi08c}. 
In our previous studies \cite{taniguchi08a,taniguchi08d,taniguchi08b}, 
we proposed that the penetration depth was evaluated 
by spin pumping effect \cite{tserkovnyak02}; 
i.e., the creation of a pure spin current by ferromagnetic resonance (FMR), 
in the F/N/F trilayers. 
The point is that 
one F layer is in resonance and acts as the spin injector 
while the other F layer is out of resonance and acts as the spin absorber. 
Thus, by measuring the dependence of 
the FMR power spectrum of the injector 
on the thickness of the absorber, 
the penetration depth of the absorber can be evaluated. 
Because microfabrication is unnecessary, 
the FMR measurement is useful for investigating the spin relaxation. 
However, the choice of the F material is restricted to those 
that can to avoid the simultaneous resonance of the two ferromagnetic layers. 
Also, because the magnitude of the magnetization, as well as the FMR frequency, 
of the absorber increases with increasing its thickness, 
the FMR spectrum of the injector and absorber sometimes overlap \cite{comment2}. 
These factors make it difficult to evaluate the penetration depth of the transverse spin current.


\begin{figure}
\centerline{\includegraphics[width=1.0\columnwidth]{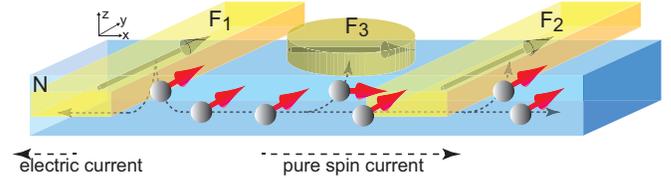}}\vspace{-3.0ex}
\caption{
         Schematic view of the nonlocal spin valve with three ferromagnetic (F) layers. 
         The black arrow in each F layer represents the magnetization. 
         The red arrow indicates the spin of the conduction electron. 
         The electric current flows from the F${}_{1}$ to the N layer 
         while the pure spin current flows to the F${}_{2}$ and F${}_{3}$ layers. 
         \vspace{-3ex}}
\label{fig:fig1}
\end{figure}



In this paper, 
we propose an alternative method to determine the penetration depth
based on the nonlocal geometry 
shown in Fig. \ref{fig:fig1}. 
The three ferromagnetic layers are attached to the nonmagnetic layer,
where the F${}_{1}$, F${}_{2}$ and F${}_{3}$ layers act as 
the injector, detector, and absorber of the spin current. 
By fabricating the three F layers in different shapes, 
we can control the directions of the magnetization in each layer independently; 
thus, a noncollinear alignment of the magnetizations can be achieved. 
For example, 
in Fig. \ref{fig:fig1}, 
the F${}_{3}$ layer is assumed to be a cylinder 
with zero in-plane shape anisotropy; 
thus, the magnetization can rotate in the plane, 
while the magnetizations of the F${}_{1}$ and F${}_{2}$ layers are parallel to the stripe 
due to the shape anisotropy. 
Let us assume that the magnetization of the F${}_{3}$ layer is perpendicular to those of the F${}_{1}$ and F${}_{3}$ layers. 
Then, the amount of the magnetoresistance measured in the F${}_{2}$ layer 
depends on the absorption of the transverse spin current in the F${}_{3}$ layer. 
Thus, by measuring the dependence of the magnetoresistance on the thickness of the F${}_{3}$ layer, 
its penetration depth can be evaluated. 
There is no restriction in the choice of the materials 
and the change of the magnitude of the magnetization does not affect the measurement, 
which are advantages compared to the FMR method. 


The paper is organized as follows. 
In Sec. \ref{sec:Spin_Current_and_Spin_Accumulation}, 
we discuss how to calculate the spin current at each F/N interface 
by solving the diffusion equation of the spin accumulation. 
Then, the amount of the magnetoresistance 
in the conventional system is quantitatively estimated in Sec. \ref{sec:Magnetoresistance}. 
In this section, 
we also show that the penetration depth of the transverse spin current can be evaluated 
by nonlocal geometry. 
Section \ref{sec:Summary} provides a summary of the work. 
The details of the calculations are shown in the Appendix. 



\section{Spin Current and Spin Accumulation}
\label{sec:Spin_Current_and_Spin_Accumulation}


In this section, 
we show the details of the calculation 
of the spin current and spin accumulation in the F and N layers, 
which are required to evaluate the magnetoresistance effect. 




\begin{figure}
\centerline{\includegraphics[width=0.8\columnwidth]{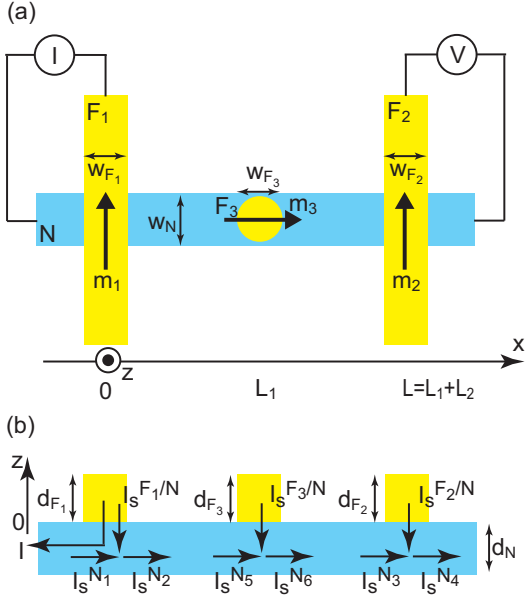}}\vspace{-3.0ex}
\caption{
         (a) The top and (b) side views of the system. 
         \vspace{-3ex}}
\label{fig:fig2}
\end{figure}



The details of the system we consider are schematically shown in Figs. \ref{fig:fig2} (a) and (b). 
The $x$ and $z$ axes are taken to be parallel and normal to the $N$ layer, 
whose origins are located at the F${}_{1}$/N interface. 
The thickness and width of the F${}_{k}$ layer are denoted as $d_{{\rm F}_{k}}$ and $w_{{\rm F}_{k}}$ 
while those of the N layer are $d_{\rm N}$ and $w_{\rm N}$. 
The distance between the F${}_{1}$ (F${}_{2}$) and the F${}_{3}$ layers are denoted as $L_{1}$ ($L_{2}$), respectively.
The unit vector pointing in the direction of the magnetization in the F${}_{k}$ layer ($k=1,2,3$) is denoted as $\mathbf{m}_{k}$. 
The electric current, $I$, flows from the F${}_{1}$ to the N layer. 
By passing through the F${}_{1}$ layer, 
the conduction electrons induce spin accumulation 
inside and the interface of the F${}_{1}$ and N layers. 
These spin accumulations are diffused in the N layer, 
creating a pure spin current into the F${}_{2}$ and F${}_{3}$ layers. 
Then, the spin accumulations are created in these F layers. 
It should be noted that the following formula is applicable 
to an arbitrary alignment of the magnetizations, $(\mathbf{m}_{1},\mathbf{m}_{2},\mathbf{m}_{3})$. 


As shown in Sec. \ref{sec:Magnetoresistance}, 
the magnetoresistance effect is determined by 
the spin current at the F/N interface. 
According to the spin dependent formula \cite{brataas01}, 
the spin current at the F/N interface 
(into N, see $\mathbf{I}_{s}^{{\rm F}_{k}/{\rm N}}$ ($k=1,2,3$) in Fig. \ref{fig:fig2}(b)) is given by 
\begin{equation}
\begin{split}
  \mathbf{I}_{s}^{\rm F/N}
  \!=\!
  \frac{1}{4\pi}\!
  &
  \left\{
    \left[
      \frac{(1 \!-\! \gamma_{\rm F/N}^{2})}{2}
      g_{\rm F/N}
      \mathbf{m}
      \!\cdot\!
      \left(
        \bm{\mu}_{\rm F}
        \!-\!
        \bm{\mu}_{\rm N}
      \right)
      \!+\!
      \frac{h}{e}
      \gamma_{\rm F/N}
      I^{\rm F/N}
    \right]\!
    \mathbf{m}
  \right.
\\
  &\!-\!
    g_{\rm r(F/N)}^{\uparrow\downarrow}
    \mathbf{m}
    \!\times\!
    \left(
      \bm{\mu}_{\rm N}
      \!\times\!
      \mathbf{m}
    \right)
    \!-\!
    g_{\rm i(F/N)}^{\uparrow\downarrow}
    \bm{\mu}_{\rm N}
    \!\times\!
    \mathbf{m}
\\
  &\!+\!
  \left.
    t_{\rm r(F/N)}^{\uparrow\downarrow}
    \mathbf{m}
    \!\times\!
    \left(
      \bm{\mu}_{\rm F}
      \!\times\!
      \mathbf{m}
    \right)
    \!+\!
    t_{\rm i}^{\uparrow\downarrow}
    \bm{\mu}_{\rm F}
    \!\times\!
    \mathbf{m}
  \right]
  \label{eq:spin_current}
\end{split}
\end{equation}
where $I^{\rm F/N}$ is the electric current from the F to N layer, 
$g_{\rm F/N}=g^{\uparrow\uparrow}+g^{\downarrow\downarrow}$ is 
the sum of the spin-up and spin-down conductance, 
$\gamma=(g^{\uparrow\uparrow}-g^{\downarrow\downarrow})/(g^{\uparrow\uparrow}+g^{\downarrow\downarrow})$ 
is the spin polarization of the interface conductance \cite{valet93}, 
$g_{\rm r(i)}^{\uparrow\downarrow}$ is the real (imaginary) part of the mixing conductance \cite{brataas01}, 
and $t_{\rm r(i)}^{\uparrow\downarrow}$ is the real (imaginary) part of the transmission mixing conductance \cite{taniguchi08a,taniguchi08b,taniguchi08c}. 
The spin accumulation in the N and F layers are denoted as $\bm{\mu}_{\rm N}$ and $\bm{\mu}_{\rm F}$, 
respectively. 
The first term ($\propto \mathbf{m}$) in Eq. (\ref{eq:spin_current}) describes the longitudinal spin transport ($\parallel \mathbf{m}$), 
while the other terms describe the transverse spin transport ($\perp \mathbf{m}$). 
The terms proportional to $g^{\uparrow\downarrow}$ describe the transverse spin injection 
from the N to F layer, 
while the terms proportional to $t^{\uparrow\downarrow}$ describe the opposite flow of the spin current. 
In the zero penetration depth limit, 
the spin accumulation in the F layer is parallel to $\mathbf{m}$, 
and the last two terms ($\propto t^{\uparrow\downarrow}$) in Eq. (\ref{eq:spin_current}) can be neglected, 
which is assumed in Ref. \cite{brataas01}. 
Although Ref. \cite{brataas01} assumes a spatially uniform spin accumulation, 
it has been shown that circuit theory is applicable to the diffusive system \cite{bauer03,tserkovnyak03}. 


The spin accumulation in the N layer obeys the diffusion equation \cite{valet93}
\begin{equation}
  \frac{d^{2}}{d x^{2}}
  \bm{\mu}_{\rm N}
  =
  \frac{1}{\lambda_{\rm N}^{2}}
  \bm{\mu}_{\rm N},
  \label{eq:diffusion_equation_N}
\end{equation}
where $\lambda_{\rm N}$ is the spin diffusion length of the N layer. 
The solution of $\bm{\mu}_{\rm N}$ is expressed as a linear combination of $e^{\pm x/\lambda_{\rm N}}$. 
The spin accumulation is related to the spin current by 
\begin{equation}
  \mathbf{I}_{s}^{\rm N}
  =
  -\frac{d}{dx}
  \frac{\hbar S_{\rm N} \sigma_{\rm N}}{2e^{2}}
  \bm{\mu}_{\rm N},
  \label{eq:spin_current_N}
\end{equation}
where $S_{\rm N}=w_{\rm N}d_{\rm N}$ and $\sigma_{\rm N}$ are 
the cross sectional area and the conductivity of the N layer, respectively. 


The longitudinal spin accumulation in the F layer,
$\bm{\mu}_{\rm F}^{\rm L}=(\mathbf{m}\cdot\bm{\mu}_{\rm F})\mathbf{m}$, also obeys 
the diffusion equation, 
\begin{equation}
  \frac{d^{2}}{dz^{2}}
  \bm{\mu}_{\rm F}^{\rm L}
  =
  \frac{1}{\lambda_{\rm F}^{2}}
  \bm{\mu}_{\rm F}^{\rm L},
  \label{eq:diffusion_equation_F_long}
\end{equation}
and its solution is expressed as a linear combination of $e^{\pm z/\lambda_{\rm F}}$, 
where $\lambda_{\rm F}$ is the spin diffusion length of the F layer. 
The longitudinal spin accumulation is related to the longitudinal spin current by 
\begin{equation}
  (\mathbf{m}\cdot\mathbf{I}_{s}^{\rm F})
  \mathbf{m}
  =
  -\frac{d}{dz}
  \frac{\hbar S_{\rm F}}{2e^{2}}
  \left(
    \sigma_{\rm F}^{\uparrow}
    \mu_{\rm F}^{\uparrow}
    -
    \sigma_{\rm F}^{\downarrow}
    \mu_{\rm F}^{\downarrow}
  \right)
  \mathbf{m},
  \label{eq:spin_current_F_long}
\end{equation}
where $S_{\rm F}=w_{\rm F}w_{\rm N}$ is the cross sectional area of the F layer \cite{comment4}. 
The electrochemical potential \cite{valet93} and the conductivity of the spin-$\sigma$ electron are denoted as 
$\mu_{\rm F}^{\sigma}$ and $\sigma_{\rm F}^{\sigma}$, respectively. 
The spin polarization of the conductivity is given by 
$\beta=(\sigma_{\rm F}^{\uparrow}-\sigma_{\rm F}^{\downarrow})/(\sigma_{\rm F}^{\uparrow}+\sigma_{\rm F}^{\downarrow})$. 
The resistivity is defined as $\rho_{\rm F}=1/\sigma_{\rm F}=1/(\sigma_{\rm F}^{\uparrow}+\sigma_{\rm F}^{\downarrow})$. 


The transverse spin accumulation in the F layer, 
$\bm{\mu}_{\rm F}^{\rm T}=\mathbf{m}\times(\bm{\mu}_{\rm F}\times\mathbf{m})$
obeys the following diffusion equation \cite{zhang02}:
\begin{equation}
  \frac{d^{2}}{dz^{2}}
  \bm{\mu}_{\rm F}^{\rm T}
  =
  \frac{1}{\lambda_{J}^{2}}
  \bm{\mu}_{\rm F}^{\rm T}
  \times
  \mathbf{m}
  +
  \frac{1}{\lambda_{\rm F(T)}^{2}}
  \bm{\mu}_{\rm F}^{\rm T}.
  \label{eq:diffusion_equation_F_trans}
\end{equation}
The coherence length $\lambda_{J}=\sqrt{(D_{\rm F}^{\uparrow}+D_{\rm F}^{\downarrow})\hbar/(2J)}$ relates 
to the spin-dependent diffusion constant $D_{\rm F}^{\sigma}$ and 
the exchange interaction $J$ between the conduction and local electrons \cite{zhang02}. 
The spin polarization of the diffusion constant is given by 
$\beta^{\prime}=(D_{\rm F}^{\uparrow}-D_{\rm F}^{\downarrow})/(D_{\rm F}^{\uparrow}+D_{\rm F}^{\downarrow})$. 
For simplicity, we assume that $\beta^{\prime}=\beta$. 
Then the transverse spin diffusion length $\lambda_{\rm F(T)}$ is given by $\lambda_{\rm F(T)}=\lambda_{\rm F}/\sqrt{1-\beta^{2}}$ \cite{zhang02}. 
The solution of Eq. (\ref{eq:diffusion_equation_F_trans}) is expressed as a 
linear combination of $e^{\pm z/\ell_{+}}$ and $e^{-\pm z/\ell_{-}}$, 
where 
\begin{equation}
  \frac{1}{\ell_{\pm}}
  =
  \sqrt{
    \frac{1}{\lambda_{\rm F(T)}^{2}}
    \mp
    \frac{i}{\lambda_{J}^{2}}
  }.
\end{equation}
The penetration depth of the transverse spin current, $\lambda_{\rm t}$, is defined as 
\begin{equation}
  \frac{1}{\lambda_{\rm t}}
  =
  {\rm Re}
  \left[
    \frac{1}{\ell_{+}}
  \right].
\end{equation}
The transverse spin current in the F layer relates to the spin accumulation as follows:
\begin{equation}
  \mathbf{m}
  \times
  \left(
    \mathbf{I}_{s}^{\rm F}
    \times
    \mathbf{m}
  \right)
  =
  -\frac{d}{dz}
  \frac{\hbar S_{\rm F} \sigma_{\rm F}^{\uparrow\downarrow}}{2e^{2}}
  \bm{\mu}_{\rm F}^{\rm T},
\end{equation}
where $\sigma_{\rm F}^{\uparrow\downarrow}=[\sigma_{\rm F}^{\uparrow}/(1+\beta^{\prime})+\sigma_{\rm F}^{\downarrow}/(1-\beta^{\prime})]/2$. 


We assume that the spin current is continuous at each F/N interface, 
and that the electric current is constant. 
Thus, the spin currents in the N layer near the interfaces, 
denoted as $\mathbf{I}_{s}^{{\rm N}_{i}}$ ($i=1-6$) in Fig. \ref{fig:fig2}(b), 
should satisfy the relation $\mathbf{I}_{s}^{{\rm N}_{2k-1}}+\mathbf{I}_{s}^{{\rm F}_{k}/{\rm N}}=\mathbf{I}_{s}^{{\rm N}_{2k}}$ 
($k=1,2,3$). 
We also assume that at the ends of the N layer, 
the spin current is zero. 
Using these boundary conditions, 
we can solve the diffusion equations of $\bm{\mu}_{\rm F}$ and $\bm{\mu}_{\rm N}$, 
where the spin current at the F/N interface can be chosen as 
the integral constant of the diffusion equation of the spin accumulation. 
Then, the spin accumulations on the right hand side of Eq. (\ref{eq:spin_current}) can be expressed 
in terms of the spin currents at each F/N interface. 
Thus, $\mathbf{I}_{s}^{\rm F/N}$ can be obtained 
by solving their simultaneous equation; 
see Appendix A.



\section{Magnetoresistance}
\label{sec:Magnetoresistance}


In this section, 
we show the details of the calculation of the magnetoresistance. 
Also, we show that the penetration depth of the transverse spin current of the F${}_{3}$ layer 
can be estimated by measuring the dependence of the magnetoresistance 
on the thickness of the F${}_{3}$ layer. 


By using the spin currents at each F/N interface 
obtained in the previous section, 
the magnetoresistance measured in the F${}_{2}$ layer is calculated as 
$R=R^{(1)}+R^{(2)}$, 
where $R^{(1)}$ and $R^{(2)}$ represent 
the contributions to the resistance from the spin dependent transports 
inside the F${}_{2}$ layer 
and the interface at the F${}_{2}$/N layers, respectively \cite{valet93,taniguchi11e}. 
The explicit forms of $R^{(1)}$ and $R^{(2)}$ are given by 
\begin{equation}
\begin{split}
  R^{(1)}
  &=
  -\frac{\beta_{\rm F_{2}}S_{\rm F_{2}}}{2eI}
  \left[
    \mathbf{m}_{2}
    \cdot
    \bm{\mu}_{\rm F_{2}}
    (z \!=\! d_{\rm F_{2}})
    -
    \mathbf{m}_{2}
    \cdot
    \bm{\mu}_{\rm F_{2}}
    (z \!=\! 0)
  \right]
\\
  &=
  -\frac{\beta_{\rm F_{2}}hS_{\rm F_{2}}}{2e^{2}g_{\rm F_{2}}}
  \tanh
  \left(
    \frac{d_{\rm F_{2}}}{2\lambda_{\rm F_{2}}}
  \right)
  \frac{2e}{\hbar I}
  \mathbf{m}_{2}
  \cdot
  \mathbf{I}_{s}^{\rm F_{2}/N},
  \label{eq:resistance_1}
\end{split}
\end{equation}
\begin{equation}
\begin{split}
  R^{(2)}
  &=
  -\frac{\gamma_{\rm F_{2}/N}}{2eI}
  \left[
    S_{\rm F_{2}}
    \mathbf{m}_{2}
    \cdot
    \bm{\mu}_{\rm F_{2}}
    (z \!=\! 0)
    -
    S_{\rm N}
    \mathbf{m}_{2}
    \cdot
    \bm{\mu}_{\rm N}
    (x \!=\! L)
  \right]
\\
  &=
  \frac{\gamma_{\rm F_{2}/N}hS_{\rm F_{2}}}{2e^{2}g_{\rm F_{2}}}
  \frac{1}{\tanh(d_{\rm F_{2}}/\lambda_{\rm F_{2}})}
  \frac{2e}{\hbar I}
  \mathbf{m}_{2}
  \cdot
  \mathbf{I}_{s}^{\rm F_{2}/N}
\\
  &\ \ \ +
  \frac{\gamma_{\rm F_{2}/N}hS_{\rm N}}{2e^{2}g_{\rm N}}
  \mathbf{m}_{2}
  \cdot
  \mathbf{I}_{s}^{\rm N{4}}, 
  \label{eq:resistance_2}
\end{split}
\end{equation}
where $g_{\rm F}/S_{\rm F}=h(1-\beta_{\rm F}^{2})/(2e^{2}\rho_{\rm F}\lambda_{\rm F})$ 
and $g_{\rm N}/S_{\rm N}=h/(2e^{2}\rho_{\rm N}\lambda_{\rm N})$, respectively. 
It should be noted that 
the above formula reproduces the results of Takahashi and Maekawa \cite{takahashi03} 
by neglecting the F${}_{3}$ layer 
and assuming that $\mathbf{m}_{1} = \pm \mathbf{m}_{2}$. 



\begin{figure}
\centerline{\includegraphics[width=0.7\columnwidth]{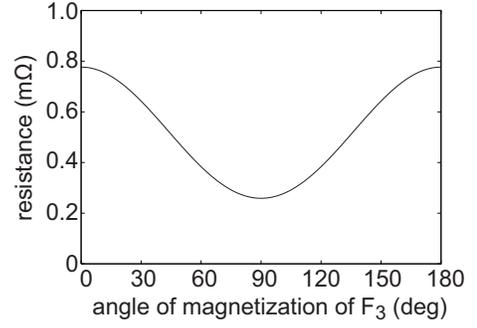}}\vspace{-3.0ex}
\caption{
         The dependence of the magnetoresistance 
         on the relative angle between $\mathbf{m}_{1}$ and $\mathbf{m}_{3}$. 
         The magnetization of the F${}_{2}$ layer is parallel to $\mathbf{m}_{1}$. 
         \vspace{-3ex}}
\label{fig:fig3}
\end{figure}



Figure \ref{fig:fig3} shows an example of the calculation result 
using Eqs. (\ref{eq:resistance_1}) and (\ref{eq:resistance_2}). 
The magnetizations of the F${}_{1}$ and F${}_{2}$ layers are assumed to be parallel ($\mathbf{m}_{1}=\mathbf{m}_{2}$) 
while the magnetization of the F${}_{3}$ layer changes its direction 
from $\mathbf{m}_{3}=\mathbf{m}_{1}$ to $\mathbf{m}_{3}=-\mathbf{m}_{1}$. 
The material parameters of the F${}_{1}$, F${}_{2}$ and F${}_{3}$ layers are assumed to be identical, 
for simplicity, 
and are taken to be 
$\rho_{\rm F}=220$ $\Omega$nm, 
$\lambda_{\rm F}=5.0$ nm,
$\lambda_{J}=2.8$ nm,
$\beta_{\rm F}=0.35$,
$\rho_{\rm N}=12$$\Omega$ nm,
$\lambda_{\rm N}=1300$ nm,
$R_{\rm F/N}=hS_{\rm F}/(e^{2}g_{\rm F/N})=275.4$ $\Omega$nm${}^{2}$,
$\gamma_{\rm F/N}=0.70$,
$g_{\rm r(F/N)}^{\uparrow\downarrow}/S_{\rm F}=38.0$ nm${}^{-2}$,
$g_{\rm i(F/N)}^{\uparrow\downarrow}/S_{\rm F}=1.0$ nm${}^{-2}$,
$t_{\rm r(F/N)}^{\uparrow\downarrow}/S_{\rm F}=4.0$ nm${}^{-2}$,
$t_{\rm i(F/N)}^{\uparrow\downarrow}/S_{\rm F}=4.0$ nm${}^{-2}$,
$d_{\rm F}=30$ nm,
$d_{\rm N}=150$ nm,
$w_{\rm F_{1}}=w_{\rm F_{2}}=120$ nm, 
$w_{\rm F_{3}}=150$ nm, 
$w_{\rm N}=150$nm, 
$L_{1}=400$ nm,
and $L_{2}=400$ nm
\cite{zhang02,taniguchi08a,comment1,fert99,bass07,comment3}. 
Then, the penetration depth is obtained as $\lambda_{\rm t}=3.5$ nm. 
The resistance area, Eqs. (\ref{eq:resistance_1}) and (\ref{eq:resistance_2}), 
is converted to resistance 
by using $S_{\rm F_{2}}=w_{\rm F_{2}}w_{\rm N}$, 
as measured in the experiment \cite{kimura06,kimura07a,kimura07b,kimura08}.


As shown in Fig. \ref{fig:fig3}, 
the resistance, as well as the amounts of spin accumulation, 
of the parallel and antiparallel $(\mathbf{m}_{3}=\pm \mathbf{m}_{1})$ alignments are 
larger than those of the perpendicular alignment $(\mathbf{m}_{3} \perp \mathbf{m}_{1})$, 
because of the fast relaxation of the transverse spin accumulation 
compared to that of the longitudinal one 
and because of the continuity of the spin accumulation. 




\begin{figure}
\centerline{\includegraphics[width=0.7\columnwidth]{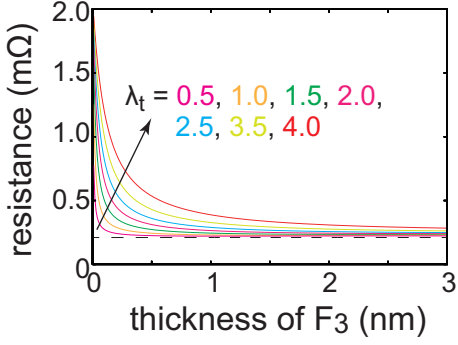}}\vspace{-3.0ex}
\caption{
         The dependences of the magnetoresistance 
         on the thickness of the F${}_{3}$ layer 
         for the various penetration depth. 
         The magnetizations of the F${}_{2}$ and F${}_{3}$ layers are assumed to be 
         $\mathbf{m}_{2} \parallel \mathbf{m}_{1}$ and 
         $\mathbf{m}_{3} \perp \mathbf{m}_{1}$, respectively. 
         The dashed line is obtained by neglecting the transverse spin accumulation. 
         \vspace{-3ex}}
\label{fig:fig4}
\end{figure}



In Fig. \ref{fig:fig4}, 
we show the dependences of the magnetoresistance 
on the thickness of the F${}_{3}$ layer, $d_{\rm F_{3}}$ 
for various penetration depths. 
The values of the parameters except $d_{\rm F_{3}}$ and $\lambda_{J}$ are the same as those in Fig. \ref{fig:fig3}. 
The penetration depth is changed by changing the value of $\lambda_{J}$, 
where $(\lambda_{J},\lambda_{\rm t})=$(0.7,1.0), (1.1,1.5), (1.5,2.0), (1.9,2.5), (2.3,3.0), (2.8,3.5), (3.5,4.0) nm, respectively. 
The magnetization alignment is assumed to be $\mathbf{m}_{2}=\mathbf{m}_{1}$ and $\mathbf{m}_{3} \perp \mathbf{m}_{1}$. 
The magnetoresistance decreases as the thickness of the F${}_{3}$ layer increases 
because the amount of the spin accumulation created around the F${}_{2}$ decreases
due to the spin absorption in the F${}_{3}$ layer. 
The magnetoresistance is constant for a sufficiently large thickness of $d_{\rm F_{3}} \gg \lambda_{\rm t}$. 
The reduction of the magnetoresistance is fast for a small $\lambda_{\rm t}$ 
because of the fast absorption of the spin accumulation. 
We also show the magnetoresistance obtained by neglecting the transverse spin accumulation. 
In this case, 
since the spin absorption in the F${}_{3}$ layer occurs only at the F${}_{3}$/N interface, 
and the amount of the spin absorption is independent of its thickness, 
the magnetoresistance is also independent of the thickness. 
In other words, 
the thickness dependence of the magnetoresistance reflects the relaxation of the transverse spin current in the F${}_{3}$ layer. 
Thus, the penetration depth of the F${}_{3}$ layer can be evaluated. 



\section{Summary}
\label{sec:Summary}

In summary, 
we show a theoretical formula to calculate the magnetoresistance 
in a nonlocal spin valve with three ferromagnetic layers. 
We show that the penetration depth of the transverse spin current can be evaluated 
by measuring the dependence of the magnetoresistance 
on the thickness of the spin absorber. 


\section*{Acknowledgement}

The authors would like to acknowledge S. Yakata and T. Kimura 
for the valuable discussions they had with us. 


\section*{Appendix A: Equations to Determine Spin Currents}

Here we show the details of the calculation of the spin currents at each F/N interface 
to evaluate Eqs. (\ref{eq:resistance_1}) and (\ref{eq:resistance_2}). 
First, since the magnetoresistance depends on 
the relative directions of the magnetizations, 
we introduce three sets of unit vectors, 
$(\mathbf{e}_{kx},\mathbf{e}_{ky},\mathbf{m}_{k})$ ($k=1,2,3$), 
which satisfy $\mathbf{e}_{kx} \times \mathbf{e}_{ky}=\mathbf{m}_{k}$. 
The rotational transformation from $(\mathbf{e}_{2x},\mathbf{e}_{2y},\mathbf{m}_{2})$ 
to $(\mathbf{e}_{1x},\mathbf{e}_{1y},\mathbf{m}_{1})$ is 
characterized by the rotation matrix $\mathsf{R}_{1}$, 
which depends on the two angle parameters $(\theta_{1},\varphi_{1})$ 
and is given by 
\begin{equation}
  \mathsf{R}_{1}
  =
  \begin{pmatrix}
    \cos\theta_{1}\cos\varphi_{1} & \cos\theta_{1}\sin\varphi_{1} & -\sin\theta_{1} \\
    -\sin\varphi_{1} & \cos\varphi_{1} & 0 \\
    \sin\theta_{1}\cos\varphi_{1} & \sin\theta_{1}\sin\varphi_{1} & \cos\theta_{1} 
  \end{pmatrix}. 
\end{equation}
Similarly, we introduce the rotation matrix $\mathsf{R}_{3}$ 
which represents the relative direction of 
$(\mathbf{e}_{3x},\mathbf{e}_{3y},\mathbf{m}_{3})$ 
with respect to $(\mathbf{e}_{2x},\mathbf{e}_{2y},\mathbf{m}_{2})$
and depends on $(\theta_{3},\varphi_{3})$. 
Although the choice of the direction of the transverse unit vectors, $\mathbf{e}_{kx}$ and $\mathbf{e}_{ky}$, is somewhat arbitrary, 
the final results is independent of these choice. 
The relation of the longitudinal and transverse spin currents between the different F layers 
can be given by $(\theta_{k},\varphi_{k})$ \cite{comment5}. 

The general solution of the diffusion equation of the longitudinal spin accumulation is given by 
\begin{equation}
\begin{split}
  \mathbf{m}
  \cdot
  \bm{\mu}_{\rm F}
  =
  \frac{4\pi}{g_{\rm F}\sinh(d_{\rm F}/\lambda_{\rm F})}
  &
  \left[
    \left(
      I_{s_{z}}^{(1)}
      +
      \frac{\hbar \beta_{\rm F}}{2e}
      I
    \right)
    \cosh
    \left(
      \frac{z-d_{\rm F}}{\lambda_{\rm F}}
    \right)
  \right.
\\
  &- 
  \left.
    \left(
      I_{s_{z}}^{(2)}
      +
      \frac{\hbar\beta_{\rm F}}{2e}
      I
    \right)
    \cosh
    \left(
      \frac{z}{\lambda_{\rm F}}
    \right)
  \right],
  \label{eq:longitudinal_spin_accumulation}
\end{split}
\end{equation}
where $I_{s_{z}}^{(1)}$ and $I_{s_{z}}^{(2)}$ are 
the longitudinal spin current at $z=0$ and $z=d_{\rm F}$, respectively, 
flowing in the positive $z$ direction. 
In the present study, as shown in Fig. \ref{fig:fig2} (b), 
$I_{s_{z}}^{(1)}=-\mathbf{m}\cdot\mathbf{I}_{s}^{\rm F/N}$ and 
$I_{s_{z}}^{(2)}=0$. 
The electric current $I$ is positive for the electron flow in the positive $z$ direction, 
and is nonzero only in the F${}_{1}$ layer. 
The general solution of the spin accumulation in the N layer 
can be obtained by replacing the quantities of the F layer in Eq. (\ref{eq:longitudinal_spin_accumulation})
with those of the N layer 
($\beta$ in the N layer is zero). 

Similarly, 
the general solution of the transverse spin accumulation is given by 
\begin{equation}
  \mathbf{e}_{x}
  \cdot
  \bm{\mu}_{\rm F}
  =
  4\pi
  \left[
    f_{1}
    I_{s_{x}}^{(1)}
    +
    i f_{2}
    I_{s_{y}}^{(1)}
    +
    f_{3}
    I_{s_{x}}^{(2)}
    +
    i f_{4}
    I_{s_{y}}^{(2)}
  \right],
  \label{eq:transverse_spin_accumulation_1}
\end{equation}
\begin{equation}
  \mathbf{e}_{y}
  \cdot
  \bm{\mu}_{\rm F}
  =
  4\pi
  \left[
    -i f_{2}
    I_{s_{x}}^{(1)}
    +
    f_{1}
    I_{s_{y}}^{(1)}
    -
    i f_{4}
    I_{s_{x}}^{(2)}
    +
    f_{3}
    I_{s_{y}}^{(2)}
  \right],
  \label{eq:transverse_spin_accumulation_2}
\end{equation}
where $f_{j}(z)$ ($j=1-4$) are given by 
\begin{equation}
  f_{1}(z)
  =
  {\rm Re}
  \left[
    \frac{\cosh[(z-d_{\rm F})/\ell_{+}]}{g_{\rm t}\sinh(d_{\rm F}/\ell_{+})}
  \right],
\end{equation}
\begin{equation}
  f_{2}(z)
  =
  i{\rm Im}
  \left[
    \frac{\cosh[(z-d_{\rm F})/\ell_{+}]}{g_{\rm t}\sinh(d_{\rm F}/\ell_{+})}
  \right],
\end{equation}
\begin{equation}
  f_{3}(z)
  =
  -{\rm Re}
  \left[
    \frac{\cosh(z/\ell_{+})}{g_{\rm t}\sinh(d_{\rm F}/\ell_{+})}
  \right],
\end{equation}
\begin{equation}
  f_{4}(z)
  =
  -i{\rm Im}
  \left[
    \frac{\cosh(z/\ell_{+})}{g_{\rm t}\sinh(d_{\rm F}/\ell_{+})}
  \right].
\end{equation}
Here $g_{\rm t}/S_{\rm F}=[h/(4e^{2}\rho_{\rm F}\ell_{+})][(1+\beta)/(1+\beta^{\prime})+(1-\beta)/(1-\beta^{\prime})]$ \cite{taniguchi08b}. 
$I_{s_{x}}^{(1)}$ and $I_{s_{y}}^{(1)}$ are the transverse spin current at $z=0$ 
while $I_{s_{x}}^{(2)}$ and $I_{s_{y}}^{(2)}$ are those at $z=d_{\rm F}$. 
In the present study, 
$I_{s_{x}}^{(1)}=-\mathbf{e}_{x}\cdot\mathbf{I}_{s}^{\rm F/N}$ and $I_{s_{y}}^{(1)}=-\mathbf{e}_{y}\cdot\mathbf{I}_{s}^{\rm F/N}$ 
while $I_{s_{x}}^{(2)}=I_{s_{y}}^{(2)}=0$. 

By using the boundary conditions mentioned in the main text 
and the solutions of the spin accumulation above, 
we obtain the following simultaneous equations of the spin currents,
\begin{equation}
  \mathsf{M}
  \mathbf{a}
  =
  \frac{\hbar I}{2e}
  \mathbf{s}. 
  \label{eq:matrix_equation}
\end{equation}
Here, $\mathsf{M}$ is the 18$^{\rm th}$ degree coefficient matrix, 
whose explicit form is given in Appendix B. 
The 18$^{\rm th}$ degree vector $\mathbf{a}$ consists of 
the spin currents at each interface, 
and is given by 
$\mathbf{a}_{1}=\mathbf{m}_{1}\cdot\mathbf{I}_{s}^{\rm F_{1}/N}$, 
$\mathbf{a}_{2}=\mathbf{e}_{1x}\cdot\mathbf{I}_{s}^{\rm F_{1}/N}$,
$\mathbf{a}_{3}=\mathbf{e}_{1y}\cdot\mathbf{I}_{s}^{\rm F_{1}/N}$,
$\mathbf{a}_{4}=\mathbf{m}_{1}\cdot\mathbf{I}_{s}^{\rm N_{2}}$, 
$\mathbf{a}_{5}=\mathbf{e}_{1x}\cdot\mathbf{I}_{s}^{\rm N_{2}}$,
$\mathbf{a}_{6}=\mathbf{e}_{1y}\cdot\mathbf{I}_{s}^{\rm N_{2}}$,
$\mathbf{a}_{7}=\mathbf{m}_{2}\cdot\mathbf{I}_{s}^{\rm F_{2}/N}$, 
$\mathbf{a}_{8}=\mathbf{e}_{2x}\cdot\mathbf{I}_{s}^{\rm F_{2}/N}$,
$\mathbf{a}_{9}=\mathbf{e}_{2y}\cdot\mathbf{I}_{s}^{\rm F_{2}/N}$,
$\mathbf{a}_{10}=\mathbf{m}_{2}\cdot\mathbf{I}_{s}^{\rm N_{4}}$, 
$\mathbf{a}_{11}=\mathbf{e}_{2x}\cdot\mathbf{I}_{s}^{\rm N_{4}}$,
$\mathbf{a}_{12}=\mathbf{e}_{2y}\cdot\mathbf{I}_{s}^{\rm N_{4}}$,
$\mathbf{a}_{13}=\mathbf{m}_{3}\cdot\mathbf{I}_{s}^{\rm F_{3}/N}$, 
$\mathbf{a}_{14}=\mathbf{e}_{3x}\cdot\mathbf{I}_{s}^{\rm F_{3}/N}$,
$\mathbf{a}_{15}=\mathbf{e}_{3y}\cdot\mathbf{I}_{s}^{\rm F_{3}/N}$,
$\mathbf{a}_{16}=\mathbf{m}_{3}\cdot\mathbf{I}_{s}^{\rm N_{6}}$, 
$\mathbf{a}_{17}=\mathbf{e}_{3x}\cdot\mathbf{I}_{s}^{\rm N_{6}}$,
$\mathbf{a}_{18}=\mathbf{e}_{3y}\cdot\mathbf{I}_{s}^{\rm N_{6}}$.
The source term of the spin current and spin accumulation is given by 
the 18$^{\rm th}$ degree vector $\mathbf{s}$,  
whose non-zero component is only $\mathbf{s}_{1}$, 
\begin{equation}
  \mathbf{s}_{1}
  =
  \gamma_{\rm F_{1}/N}
  +
  \beta_{\rm F_{1}}
  \frac{(1-\gamma_{\rm F_{1}/N}^{2})g_{\rm F_{1}/N}}{2g_{\rm F_{1}}}
  \tanh
  \left(
    \frac{d_{\rm F_{1}}}{2 \lambda_{\rm F_{1}}}
  \right).
\end{equation}
Then, by numerically calculating the inverse of $\mathsf{M}$, 
the spin currents at each F/N interface, $\mathbf{a}$, can be obtained. 


\section*{Appendix B: Explicit Form of Coefficient Matrix $\mathsf{M}$}

Here we show the explicit form of the non-zero components of the coefficient matrix $\mathsf{M}$, 
\begin{equation}
  \mathsf{M}_{1,1}
  \!=\!
  1
  \!+\!
  \frac{(1-\gamma_{\rm F_{1}/N}^{2})g_{\rm F_{1}/N}}{2g_{\rm F_{1}}\tanh(d_{\rm F_{1}}/\lambda_{\rm F_{1}})}
  \!+\!
  \frac{(1-\gamma_{\rm F_{1}/N}^{2})g_{\rm F_{1}/N}}{2g_{\rm N}},
\end{equation}
\begin{equation}
  \mathsf{M}_{1,4}
  \!=\!
  -\frac{(1-\gamma_{\rm F_{1}/N}^{2})g_{\rm F_{1}/N}}{2g_{\rm N}},
\end{equation}
\begin{equation}
\begin{split}
  \mathsf{M}_{2,2}
  \!=\!
  \mathsf{M}_{3,3}
  \!=\!&
  1
  \!+\!
  \frac{g_{\rm r(F_{1}/N)}^{\uparrow\downarrow}}{g_{\rm N}}
  \!+\!
  t_{\rm r(F_{1}/N)}^{\uparrow\downarrow}
  {\rm Re}
  \left[
    \frac{1}{g_{\rm t(F_{1})}\tanh(d_{\rm F_{1}}/\ell_{\rm F_{1}})}
  \right]
\\
  &\!+\!
  t_{\rm i(F_{1}/N)}^{\uparrow\downarrow}
  {\rm Im}
  \left[
    \frac{1}{g_{\rm t(F_{1})}\tanh(d_{\rm F_{1}}/\ell_{\rm F_{1}})}
  \right],
\end{split}
\end{equation}
\begin{equation}
\begin{split}
  \mathsf{M}_{2,3}
  \!=\!
  -\mathsf{M}_{3,2}
  =&
  \frac{g_{\rm i(F_{1}/N)}^{\uparrow\downarrow}}{g_{\rm N}}
  \!-\!
  t_{\rm r(F_{1}/N)}^{\uparrow\downarrow}
  {\rm Im}
  \left[
    \frac{1}{g_{\rm t(F_{1})}\tanh(d_{\rm F_{1}}/\ell_{\rm F_{1}})}
  \right]
\\
  &\!+\!
  t_{\rm i(F_{1}/N)}^{\uparrow\downarrow}
  {\rm Re}
  \left[
    \frac{1}{g_{\rm t(F_{1})}\tanh(d_{\rm F_{1}}/\ell_{\rm F_{1}})}
  \right],
\end{split}
\end{equation}
\begin{equation}
  \mathsf{M}_{2,5}
  \!=\!
  \mathsf{M}_{3,6}
  \!=\!
  -\frac{g_{\rm r(F_{1}/N)}^{\uparrow\downarrow}}{g_{\rm N}},
\end{equation}
\begin{equation}
  \mathsf{M}_{2,6}
  \!=\!
  -\mathsf{M}_{3,5}
  \!=\!
  -\frac{g_{\rm i(F_{1}/N)}^{\uparrow\downarrow}}{g_{\rm N}},
\end{equation}
\begin{equation}
  \mathsf{M}_{4,1}
  \!=\!
  \mathsf{M}_{5,2}
  \!=\!
  \mathsf{M}_{6,3}
  =
  1,
\end{equation}
\begin{equation}
  \mathsf{M}_{4,4}
  \!=\!
  \mathsf{M}_{5,5}
  \!=\!
  \mathsf{M}_{6,6}
  \!=\!
  -\left[
    1
    \!+\!
    \frac{1}{\tanh(L_{1}/\lambda_{\rm N})}
  \right],
\end{equation}
\begin{equation}
  \mathsf{M}_{4,13}
  \!=\!
  -\mathsf{M}_{4,16}
  \!=\!
  -\frac{1}{\sinh(L_{1}/\lambda_{\rm N})}
  (\mathsf{R}_{1}\mathsf{R}_{3}^{-1})_{33},
\end{equation}
\begin{equation}
  \mathsf{M}_{4,14}
  \!=\!
  -\mathsf{M}_{4,17}
  \!=\!
  -\frac{1}{\sinh(L_{1}/\lambda_{\rm N})}
  (\mathsf{R}_{1}\mathsf{R}_{3}^{-1})_{31},
\end{equation}
\begin{equation}
  \mathsf{M}_{4,15}
  \!=\!
  -\mathsf{M}_{4,18}
  \!=\!
  -\frac{1}{\sinh(L_{1}/\lambda_{\rm N})}
  (\mathsf{R}_{1}\mathsf{R}_{3}^{-1})_{32},
\end{equation}
\begin{equation}
  \mathsf{M}_{5,13}
  \!=\!
  -\mathsf{M}_{5,16}
  \!=\!
  -\frac{1}{\sinh(L_{1}/\lambda_{\rm N})}
  (\mathsf{R}_{1}\mathsf{R}_{3}^{-1})_{13},
\end{equation}
\begin{equation}
  \mathsf{M}_{5,14}
  \!=\!
  -\mathsf{M}_{5,17}
  \!=\!
  -\frac{1}{\sinh(L_{1}/\lambda_{\rm N})}
  (\mathsf{R}_{1}\mathsf{R}_{3}^{-1})_{11},
\end{equation}
\begin{equation}
  \mathsf{M}_{5,15}
  \!=\!
  -\mathsf{M}_{5,18}
  \!=\!
  -\frac{1}{\sinh(L_{1}/\lambda_{\rm N})}
  (\mathsf{R}_{1}\mathsf{R}_{3}^{-1})_{12},
\end{equation}
\begin{equation}
  \mathsf{M}_{6,13}
  \!=\!
  -\mathsf{M}_{6,16}
  \!=\!
  -\frac{1}{\sinh(L_{1}/\lambda_{\rm N})}
  (\mathsf{R}_{1}\mathsf{R}_{3}^{-1})_{23},
\end{equation}
\begin{equation}
  \mathsf{M}_{6,14}
  \!=\!
  -\mathsf{M}_{6,17}
  \!=\!
  -\frac{1}{\sinh(L_{1}/\lambda_{\rm N})}
  (\mathsf{R}_{1}\mathsf{R}_{3}^{-1})_{21},
\end{equation}
\begin{equation}
  \mathsf{M}_{6,15}
  \!=\!
  -\mathsf{M}_{6,18}
  \!=\!
  -\frac{1}{\sinh(L_{1}/\lambda_{\rm N})}
  (\mathsf{R}_{1}\mathsf{R}_{3}^{-1})_{22},
\end{equation}
\begin{equation}
  \mathsf{M}_{7,7}
  \!=\!
  1
  \!+\!
  \frac{(1-\gamma_{\rm F_{2}/N}^{2})g_{\rm F_{2}/N}}{2g_{\rm F_{2}}\tanh(d_{\rm F_{2}}/\lambda_{\rm F_{2}})},
\end{equation}
\begin{equation}
  \mathsf{M}_{7,10}
  \!=\!
  \frac{(1-\gamma_{\rm F_{2}/N}^{2})g_{\rm F_{2}/N}}{2g_{\rm N}},
\end{equation}
\begin{equation}
\begin{split}
  \mathsf{M}_{8,8}
  \!=\!
  \mathsf{M}_{9,9}
  \!=\!&
  1
  \!+\!
  t_{\rm r(F_{2}/N)}^{\uparrow\downarrow}
  {\rm Re}
  \left[
    \frac{1}{g_{\rm t(F_{2})}\tanh(d_{\rm F_{2}}/\ell_{\rm F_{2}})}
  \right]
\\
  &\!+\!
  t_{\rm i(F_{2}/N)}^{\uparrow\downarrow}
  {\rm Im}
  \left[
    \frac{1}{g_{\rm t(F_{2})}\tanh(d_{\rm F_{2}}/\ell_{\rm F_{2}})}
  \right],
\end{split}
\end{equation}
\begin{equation}
\begin{split}
  \mathsf{M}_{8,9}
  \!=\!
  -\mathsf{M}_{9,8}
  \!=\!&
  -t_{\rm r(F_{2}/N)}^{\uparrow\downarrow}
  {\rm Im}
  \left[
    \frac{1}{g_{\rm t(F_{2})}\tanh(d_{\rm F_{2}}/\ell_{\rm F_{2}})}
  \right]
\\
  &\!+\!
  t_{\rm i(F_{2}/N)}^{\uparrow\downarrow}
  {\rm Re}
  \left[
    \frac{1}{g_{\rm t(F_{2})}\tanh(d_{\rm F_{2}}/\ell_{\rm F_{2}})}
  \right],
\end{split}
\end{equation}
\begin{equation}
  \mathsf{M}_{8,11}
  \!=\!
  \mathsf{M}_{9,12}
  \!=\!
  \frac{g_{\rm r(F_{2}/N)}^{\uparrow\downarrow}}{g_{\rm N}},
\end{equation}
\begin{equation}
  \mathsf{M}_{8,12}
  \!=\!
  -\mathsf{M}_{9,11}
  \!=\!
  \frac{g_{\rm i(F_{2}/N)}^{\uparrow\downarrow}}{g_{\rm N}},
\end{equation}
\begin{equation}
  \mathsf{M}_{10,7}
  \!=\!
  \mathsf{M}_{11,8}
  \!=\!
  \mathsf{M}_{12,9}
  \!=\!
  \frac{1}{\tanh(L_{2}/\lambda_{\rm N})},
\end{equation}
\begin{equation}
  \mathsf{M}_{10,10}
  \!=\!
  \mathsf{M}_{11,11}
  \!=\!
  \mathsf{M}_{12,12}
  =
  -\left[
    1
    \!+\!
    \frac{1}{\tanh(L_{2}/\lambda_{\rm N})}
  \right],
\end{equation}
\begin{equation}
  \mathsf{M}_{10,16}
  \!=\!
  \frac{1}{\sinh(L_{2}/\lambda_{\rm N})}
  (\mathsf{R}_{3}^{-1})_{33},
\end{equation}
\begin{equation}
  \mathsf{M}_{10,17}
  \!=\!
  \frac{1}{\sinh(L_{2}/\lambda_{\rm N})}
  (\mathsf{R}_{3}^{-1})_{31},
\end{equation}
\begin{equation}
  \mathsf{M}_{10,18}
  \!=\!
  \frac{1}{\sinh(L_{2}/\lambda_{\rm N})}
  (\mathsf{R}_{3}^{-1})_{32},
\end{equation}
\begin{equation}
  \mathsf{M}_{11,16}
  \!=\!
  \frac{1}{\sinh(L_{2}/\lambda_{\rm N})}
  (\mathsf{R}_{3}^{-1})_{13},
\end{equation}
\begin{equation}
  \mathsf{M}_{11,17}
  \!=\!
  \frac{1}{\sinh(L_{2}/\lambda_{\rm N})}
  (\mathsf{R}_{3}^{-1})_{11},
\end{equation}
\begin{equation}
  \mathsf{M}_{11,18}
  \!=\!
  \frac{1}{\sinh(L_{2}/\lambda_{\rm N})}
  (\mathsf{R}_{3}^{-1})_{12},
\end{equation}
\begin{equation}
  \mathsf{M}_{12,16}
  \!=\!
  \frac{1}{\sinh(L_{2}/\lambda_{\rm N})}
  (\mathsf{R}_{3}^{-1})_{23},
\end{equation}
\begin{equation}
  \mathsf{M}_{12,17}
  \!=\!
  \frac{1}{\sinh(L_{2}/\lambda_{\rm N})}
  (\mathsf{R}_{3}^{-1})_{21},
\end{equation}
\begin{equation}
  \mathsf{M}_{12,18}
  \!=\!
  \frac{1}{\sinh(L_{2}/\lambda_{\rm N})}
  (\mathsf{R}_{3}^{-1})_{22},
\end{equation}
\begin{equation}
  \mathsf{M}_{13,4}
  \!=\!
  \frac{(1-\gamma_{\rm F_{3}/N}^{2})g_{\rm F_{3}/N}}{2g_{\rm N}\sinh(L_{1}/\lambda_{\rm N})}
  (\mathsf{R}_{3}\mathsf{R}_{1}^{-1})_{33},
\end{equation}
\begin{equation}
  \mathsf{M}_{13,5}
  \!=\!
  \frac{(1-\gamma_{\rm F_{3}/N}^{2})g_{\rm F_{3}/N}}{2g_{\rm N}\sinh(L_{1}/\lambda_{\rm N})}
  (\mathsf{R}_{3}\mathsf{R}_{1}^{-1})_{31},
\end{equation}
\begin{equation}
  \mathsf{M}_{13,6}
  \!=\!
  \frac{(1-\gamma_{\rm F_{3}/N}^{2})g_{\rm F_{3}/N}}{2g_{\rm N}\sinh(L_{1}/\lambda_{\rm N})}
  (\mathsf{R}_{3}\mathsf{R}_{1}^{-1})_{32},
\end{equation}
\begin{equation}
  \mathsf{M}_{13,13}
  \!=\!
  1
  \!+\!
  \frac{(1-\gamma_{\rm F_{3}/N}^{2})g_{\rm F_{3}/N}}{2g_{\rm F_{3}}\tanh(d_{\rm F_{3}}/\lambda_{\rm F_{3}})}
  +
  \frac{(1-\gamma_{\rm F_{3}/N}^{2})g_{\rm F_{3}/N}}{2g_{\rm N}\tanh(L_{1}/\lambda_{\rm N})},
\end{equation}
\begin{equation}
  \mathsf{M}_{13,16}
  \!=\!
  -\frac{(1-\gamma_{\rm F_{3}/N}^{2})g_{\rm F_{3}/N}}{2g_{\rm N}\tanh(L_{1}/\lambda_{\rm N})},
\end{equation}
\begin{equation}
  \mathsf{M}_{14,4}
  \!=\!
  \frac{g_{\rm r(F_{3}/N)}^{\uparrow\downarrow}}{g_{\rm N}\sinh(L_{1}/\lambda_{\rm N})}
  (\mathsf{R}_{3}\mathsf{R}_{1}^{-1})_{13}
  +
  \frac{g_{\rm i(F_{3}/N)}^{\uparrow\downarrow}}{g_{\rm N}\sinh(L_{1}/\lambda_{\rm N})}
  (\mathsf{R}_{3}\mathsf{R}_{1}^{-1})_{23},
\end{equation}
\begin{equation}
  \mathsf{M}_{14,5}
  \!=\!
  \frac{g_{\rm r(F_{3}/N)}^{\uparrow\downarrow}}{g_{\rm N}\sinh(L_{1}/\lambda_{\rm N})}
  (\mathsf{R}_{3}\mathsf{R}_{1}^{-1})_{11}
  +
  \frac{g_{\rm i(F_{3}/N)}^{\uparrow\downarrow}}{g_{\rm N}\sinh(L_{1}/\lambda_{\rm N})}
  (\mathsf{R}_{3}\mathsf{R}_{1}^{-1})_{21}
\end{equation}
\begin{equation}
  \mathsf{M}_{14,6}
  \!=\!
  \frac{g_{\rm r(F_{3}/N)}^{\uparrow\downarrow}}{g_{\rm N}\sinh(L_{1}/\lambda_{\rm N})}
  (\mathsf{R}_{3}\mathsf{R}_{1}^{-1})_{12}
  +
  \frac{g_{\rm i(F_{3}/N)}^{\uparrow\downarrow}}{g_{\rm N}\sinh(L_{1}/\lambda_{\rm N})}
  (\mathsf{R}_{3}\mathsf{R}_{1}^{-1})_{22},
\end{equation}
\begin{equation}
  \mathsf{M}_{15,4}
  \!=\!
  \frac{g_{\rm r(F_{3}/N)}^{\uparrow\downarrow}}{g_{\rm N}\sinh(L_{1}/\lambda_{\rm N})}
  (\mathsf{R}_{3}\mathsf{R}_{1}^{-1})_{23}
  -
  \frac{g_{\rm i(F_{3}/N)}^{\uparrow\downarrow}}{g_{\rm N}\sinh(L_{1}/\lambda_{\rm N})}
  (\mathsf{R}_{3}\mathsf{R}_{1}^{-1})_{13},
\end{equation}
\begin{equation}
  \mathsf{M}_{15,5}
  \!=\!
  \frac{g_{\rm r(F_{3}/N)}^{\uparrow\downarrow}}{g_{\rm N}\sinh(L_{1}/\lambda_{\rm N})}
  (\mathsf{R}_{3}\mathsf{R}_{1}^{-1})_{21}
  -
  \frac{g_{\rm i(F_{3}/N)}^{\uparrow\downarrow}}{g_{\rm N}\sinh(L_{1}/\lambda_{\rm N})}
  (\mathsf{R}_{3}\mathsf{R}_{1}^{-1})_{11},
\end{equation}
\begin{equation}
  \mathsf{M}_{15,6}
  \!=\!
  \frac{g_{\rm r(F_{3}/N)}^{\uparrow\downarrow}}{g_{\rm N}\sinh(L_{1}/\lambda_{\rm N})}
  (\mathsf{R}_{3}\mathsf{R}_{1}^{-1})_{22}
  -
  \frac{g_{\rm i(F_{3}/N)}^{\uparrow\downarrow}}{g_{\rm N}\sinh(L_{1}/\lambda_{\rm N})}
  (\mathsf{R}_{3}\mathsf{R}_{1}^{-1})_{12},
\end{equation}
\begin{equation}
\begin{split}
  \mathsf{M}_{14,14}
  \!=\!
  \mathsf{M}_{15,15}
  \!=\!&
  1
  \!+\!
  \frac{g_{\rm r(F_{3}/N)}^{\uparrow\downarrow}}{g_{\rm N}\tanh(L_{1}/\lambda_{\rm N})}
\\
  &\!+\!
  t_{\rm r(F_{3}/N)}^{\uparrow\downarrow}
  {\rm Re}
  \left[
    \frac{1}{g_{\rm t(F_{3})}\tanh(d_{\rm F_{3}}/\ell_{\rm F_{3}})}
  \right]
\\
  &\!+\!
  t_{\rm i(F_{3}/N)}^{\uparrow\downarrow}
  {\rm Im}
  \left[
    \frac{1}{g_{\rm t(F_{3})}\tanh(d_{\rm F_{3}}/\ell_{\rm F_{3}})}
  \right],
\end{split}
\end{equation}
\begin{equation}
\begin{split}
  \mathsf{M}_{14,15}
  \!=\!
  -\mathsf{M}_{15,14}
  \!=\!&
  \frac{g_{\rm i(F_{3}/N)}^{\uparrow\downarrow}}{g_{\rm N}\tanh(L_{1}/\lambda_{\rm N})}
\\
  &\!-\!
  t_{\rm r(F_{3}/N)}^{\uparrow\downarrow}
  {\rm Im}
  \left[
    \frac{1}{g_{\rm t(F_{3})}\tanh(d_{\rm F_{3}}/\ell_{\rm F_{3}})}
  \right]
\\
  &\!+\!
  t_{\rm i(F_{3}/N)}^{\uparrow\downarrow}
  {\rm Re}
  \left[
    \frac{1}{g_{\rm t(F_{3})}\tanh(d_{\rm F_{3}}/\ell_{\rm F_{3}})}
  \right],
\end{split}
\end{equation}
\begin{equation}
  \mathsf{M}_{14,17}
  \!=\!
  \mathsf{M}_{15,18}
  \!=\!
  -\frac{g_{\rm r(F_{3}/N)}^{\uparrow\downarrow}}{g_{\rm N}\tanh(L_{1}/\lambda_{\rm N})},
\end{equation}
\begin{equation}
  \mathsf{M}_{14,18}
  \!=\!
  -\mathsf{M}_{15,17}
  \!=\!
  -\frac{g_{\rm i(F_{3}/N)}^{\uparrow\downarrow}}{g_{\rm N}\tanh(L_{1}/\lambda_{\rm N})},
\end{equation}
\begin{equation}
  \mathsf{M}_{16,4}
  \!=\!
  \frac{1}{\sinh(L_{1}/\lambda_{\rm N})}
  (\mathsf{R}_{3}\mathsf{R}_{1}^{-1})_{33},
\end{equation}
\begin{equation}
  \mathsf{M}_{16,5}
  \!=\!
  \frac{1}{\sinh(L_{1}/\lambda_{\rm N})}
  (\mathsf{R}_{3}\mathsf{R}_{1}^{-1})_{31},
\end{equation}
\begin{equation}
  \mathsf{M}_{16,6}
  \!=\!
  \frac{1}{\sinh(L_{1}/\lambda_{\rm N})}
  (\mathsf{R}_{3}\mathsf{R}_{1}^{-1})_{32},
\end{equation}
\begin{equation}
  \mathsf{M}_{17,4}
  \!=\!
  \frac{1}{\sinh(L_{1}/\lambda_{\rm N})}
  (\mathsf{R}_{3}\mathsf{R}_{1}^{-1})_{13},
\end{equation}
\begin{equation}
  \mathsf{M}_{17,5}
  \!=\!
  \frac{1}{\sinh(L_{1}/\lambda_{\rm N})}
  (\mathsf{R}_{3}\mathsf{R}_{1}^{-1})_{11},
\end{equation}
\begin{equation}
  \mathsf{M}_{17,6}
  \!=\!
  \frac{1}{\sinh(L_{1}/\lambda_{\rm N})}
  (\mathsf{R}_{3}\mathsf{R}_{1}^{-1})_{12},
\end{equation}
\begin{equation}
  \mathsf{M}_{18,4}
  \!=\!
  \frac{1}{\sinh(L_{1}/\lambda_{\rm N})}
  (\mathsf{R}_{3}\mathsf{R}_{1}^{-1})_{23},
\end{equation}
\begin{equation}
  \mathsf{M}_{18,5}
  \!=\!
  \frac{1}{\sinh(L_{1}/\lambda_{\rm N})}
  (\mathsf{R}_{3}\mathsf{R}_{1}^{-1})_{21},
\end{equation}
\begin{equation}
  \mathsf{M}_{18,6}
  \!=\!
  \frac{1}{\sinh(L_{1}/\lambda_{\rm N})}
  (\mathsf{R}_{3}\mathsf{R}_{1}^{-1})_{22},
\end{equation}
\begin{equation}
  \mathsf{M}_{16,7}
  \!=\!
  -\mathsf{M}_{16,10}
  \!=\!
  -\frac{1}{\sinh(L_{1}/\lambda_{\rm N})}
  (\mathsf{R}_{3})_{33},
\end{equation}
\begin{equation}
  \mathsf{M}_{16,8}
  \!=\!
  -\mathsf{M}_{16,11}
  \!=\!
  -\frac{1}{\sinh(L_{1}/\lambda_{\rm N})}
  (\mathsf{R}_{3})_{31},
\end{equation}
\begin{equation}
  \mathsf{M}_{16,9}
  \!=\!
  -\mathsf{M}_{16,12}
  \!=\!
  -\frac{1}{\sinh(L_{1}/\lambda_{\rm N})}
  (\mathsf{R}_{3})_{32},
\end{equation}
\begin{equation}
  \mathsf{M}_{17,7}
  \!=\!
  -\mathsf{M}_{17,10}
  \!=\!
  -\frac{1}{\sinh(L_{1}/\lambda_{\rm N})}
  (\mathsf{R}_{3})_{13},
\end{equation}
\begin{equation}
  \mathsf{M}_{17,8}
  \!=\!
  -\mathsf{M}_{17,11}
  \!=\!
  -\frac{1}{\sinh(L_{1}/\lambda_{\rm N})}
  (\mathsf{R}_{3})_{11},
\end{equation}
\begin{equation}
  \mathsf{M}_{17,9}
  \!=\!
  -\mathsf{M}_{17,12}
  \!=\!
  -\frac{1}{\sinh(L_{1}/\lambda_{\rm N})}
  (\mathsf{R}_{3})_{12},
\end{equation}
\begin{equation}
  \mathsf{M}_{18,7}
  \!=\!
  -\mathsf{M}_{18,10}
  \!=\!
  -\frac{1}{\sinh(L_{1}/\lambda_{\rm N})}
  (\mathsf{R}_{3})_{23},
\end{equation}
\begin{equation}
  \mathsf{M}_{18,8}
  \!=\!
  -\mathsf{M}_{18,11}
  \!=\!
  -\frac{1}{\sinh(L_{1}/\lambda_{\rm N})}
  (\mathsf{R}_{3})_{21},
\end{equation}
\begin{equation}
  \mathsf{M}_{18,9}
  \!=\!
  -\mathsf{M}_{18,12}
  \!=\!
  -\frac{1}{\sinh(L_{1}/\lambda_{\rm N})}
  (\mathsf{R}_{3})_{22},
\end{equation}
\begin{equation}
  \mathsf{M}_{16,13}
  \!=\!
  \mathsf{M}_{17,14}
  \!=\!
  \mathsf{M}_{18,15}
  \!=\!
  \frac{1}{\tanh(L_{1}/\lambda_{\rm N})},
\end{equation}
\begin{equation}
  \mathsf{M}_{16,16}
  \!=\!
  \mathsf{M}_{17,17}
  \!=\!
  \mathsf{M}_{18,18}
  \!=\!
  -\left[
    \frac{1}{\tanh(L_{1}/\lambda_{\rm N})}
    \!+\!
    \frac{1}{\tanh(L_{2}/\lambda_{\rm N})}
  \right].
\end{equation}


\begin{thebibliography}{36}
\bibitem[{com({\natexlab{b}})}]{comment4}
\bibinfo{note}{Although the cross sectional area of the F${}_{3}$ layer,
  $S_{\rm F_{3}}$, in Fig. \ref{fig:fig2} (a) does not equal $w_{\rm
  F_{3}}w_{\rm N}$, this does not affect the final result because the value of
  the conductance (or resistance) evaluated in both theory and experiment is
  that for per unit area.}
\bibitem[{com({\natexlab{d}})}]{comment3}
\bibinfo{note}{The values of $g_{\rm F/N}$ and $g_{\rm r}^{\uparrow\downarrow}$
  are obtained by different experiments. Although the conductance estimated by
  the scattering matrix should satisfy $2g_{\rm r}^{\uparrow\downarrow}>g_{\rm
  F/N}$ \cite{brataas01}, the measured values do not need to satify this
  relation because of the effect of the Sharvin conductance \cite{xia01}.}
\bibitem[{com({\natexlab{e}})}]{comment5}
\bibinfo{note}{In a system consisting of two ferromagnetic layers, the
  magnetoresistance depends on one angle (the relative angle between the
  magnetizations). In the present case, we can choose the system in spin space
  with $\varphi_{1}=0$ without losing generality. Thus, strictly, the
  magnetoresistance depends on the three angles ($\theta_{1}$, $\theta_{2}$,
  and $\varphi_{2}$). However, here we keep four angles to make the expressions
  of $\mathsf{R}_{1}$ and $\mathsf{R}_{3}$ identical.}
\end{thebibliography}
\end{document}